\documentclass[preprint,12pt,sort&compress]{elsarticle}

\usepackage[utf8]{inputenc}
\usepackage[english]{babel}

\usepackage{setspace}
\usepackage{booktabs}
\usepackage{multirow}
\usepackage{amsmath,amssymb}
\usepackage{qcircuit}
\usepackage{lipsum}
\usepackage{natbib}
\usepackage{stfloats}
\usepackage{graphicx}
\usepackage{caption} 
\usepackage{subcaption} 

\usepackage{physics}
\usepackage[figurename=Fig., justification=RaggedRight]{caption}
\usepackage{amsmath}
\usepackage{xcolor}
\usepackage{hyperref}

\begin{document}

\title{Grover's search meets Ising models: a quantum algorithm for finding low-energy states}

\author[1]{A. A. Zhukov\corref{cor1} }
\ead{zugazoid@gmail.com}
\author[1,2]{A. V. Lebedev}
\author[1,2,3]{W. V. Pogosov}
\cortext[cor1]{Corresponding author}

\affiliation[1]{organization={Dukhov Research Institute of Automatics (VNIIA)},
postcode={127030},
city={Moscow},
country={Russia}}
\affiliation[2]{organization={Advanced Mesoscience and Nanotechnology Centre, Moscow Institute of Physics and Technology (MIPT)},
city={Dolgoprudny},
postcode={141700},
country={Russia}}
\affiliation[3]{organization={Institute for Theoretical and Applied Electrodynamics, Russian Academy of Sciences},
city={Moscow},
postcode={125412},
country={Russia}}


\begin{abstract}

We propose a methodology for implementing Grover's algorithm in the digital quantum simulation of disordered Ising models. The core concept revolves around using the evolution operator for the Ising model as the quantum oracle within Grover's search. This operator induces phase shifts for the eigenstates of the Ising Hamiltonian, with the most pronounced shifts occurring for the lowest and highest energy states. Determining these states for a disordered Ising Hamiltonian using classical methods presents an exponentially complex challenge with respect to the number of spins (or qubits) involved. Within our proposed approach, we determine the optimal evolution time by ensuring a phase flip for the target states. This method yields a quadratic speedup compared to classical computation methods and enables the identification of the lowest and highest energy states (or neighboring states) with a high probability~$\lesssim 1$.

\end{abstract}

\begin{keyword}
quantum circuit \sep quantum simulation \sep Grover's search \sep Ising model \sep optimization
\end{keyword}

\maketitle

\section{INTRODUCTION}\label{s:intro}

In recent years, significant advancements have been made in the development of real quantum computers using various physical platforms \cite{Postler2022,kim2023evidence,Bluvstein2023,2023}. In particular, these platforms include trapped ions, superconducting circuits, and neutral atoms. Among the promising applications of quantum computing, quantum simulation stands out due to its potential to model complex quantum systems that are intractable for classical computers \cite{manin1980computable,FEYNMAN}. Quantum computing offers advantages such as an exponentially larger Hilbert space compared to classical systems and the ability to utilize quantum phenomena like quantum interference and parallelism \cite{nielsen2010quantum}. These features are also instrumental in achieving exponential or polynomial speedups in quantum algorithms like Shor's and Grover's algorithms, which solve problems such as integer factorization and unstructured search much faster than their classical counterparts \cite{shor1999polynomial,grover1996fast}. For recent advancements in Grover's algorithm, see, e.g., Refs. \cite{1,2,3,4,5}.

This paper aims to integrate Grover's search with quantum simulation of classical spin models such as the Ising model. It is well-known that many NP-complete and NP-hard problems can be formulated in terms of Ising models \cite{Lucas2014, Mohseni2022}, highlighting not only the fundamental but also the practical significance of these models. By leveraging the ability of quantum computers to efficiently simulate the dynamics of these models, we can potentially find solutions to these hard problems more effectively. A crucial element of Grover's algorithm is the quantum oracle, which alters the phase of the target state. We propose an analogy between this oracle and the evolution operator of the Ising Hamiltonian, which similarly modifies the phases of eigenstates in the uniform superposition depending on their energies and time. By identifying the optimal evolution time that flips the phases of the lowest or highest energy states, we can enhance the probability of finding these states in the superposition, akin to Grover's search. Additionally, we can use another evolution time to amplify probability amplitudes around arbitrary energies to target corresponding quantum states.

As a case study, we examine a disordered classical Ising model with all-to-all interactions and random energy constants. Our algorithm effectively operates as Grover's search, enabling the identification of target eigenstates or those closely related to them in energy with high probability ($\lesssim 1$). Consequently, our approach offers a quadratic speedup compared to classical brute-force search and may present a viable alternative to other quantum simulation methods for Ising models, such as quantum annealing \cite{PhysRevE.58.5355,Santoro2002,Barends2016} or variational quantum algorithms \cite{Cerezo2021,Tilly2022}. Notably, quantum annealing leverages quantum tunneling to escape local minima, while variational quantum algorithms use a hybrid quantum-classical approach to optimize a parameterized quantum circuit. Both methods have their strengths, but our approach provides a direct application of Grover's algorithm to quantum simulation, which may be particularly advantageous in the era of fault-tolerant quantum computing.  Recently, a quantum-inspired approach has been proposed, demonstrating Grover’s search on classical hardware with a logarithmic number of oracle calls \cite{stoudenmire2024opening}. However, this method assumes a low-entanglement oracle, limiting its applicability to weakly interacting systems. In contrast, the Ising model in this work features all-to-all interactions and strong entanglement, making it a more suitable candidate for quantum speedup. This reliance on quantum entanglement also means that the implementation of Grover’s algorithm is particularly sensitive to gate errors \cite{PhysRevA.102.042609}, highlighting the importance of advancements in quantum error correction and fault-tolerance.

The paper is structured as follows. In Section II, we formulate our method. Section III presents our main results. We conclude in Section IV.

\section{Method}\label{s:alg}

\subsection{Hamiltonian}

In our scenario, spins can be naturally represented by qubits. The Hamiltonian of the classical Ising model is expressed as follows:
\begin{eqnarray}
H_{\text{Ising}} = \sum_{j=1}^{N_q} \varepsilon_j \sigma_j^z + \sum_{i\neq j}^{N_q} J_{ij} \sigma_i^z \sigma_j^z,
\label{Ising}
\end{eqnarray}
where $\sigma_j^z$ denotes the $Z$-Pauli operator acting on the $j$-th spin (qubit), and $N_q$ represents the number of qubits. The total number of eigenstates of the Ising Hamiltonian is given by $N_s = 2^{N_q}$. We assume that the energy constants follow a random distribution, typically a Gaussian distribution. Here, we also assume $\bar{\varepsilon} = \bar{J} = 0$, where the averaging is carried out over the qubits, and the standard deviations for $J_{ij}$ and $\varepsilon_j$ are denoted by $\sigma_J = \sigma_{\varepsilon} = 2$.

It is worth noting that our methodology extends naturally to more complex Ising model variants \cite{Bybee2023-gx}, which incorporate higher-order terms such as $\sim \sigma_i^z \sigma_j^z \sigma_k^z$, and so forth.

\subsection{Algorithm}

\begin{figure}[h]
\centering
\includegraphics[width=0.8\textwidth]{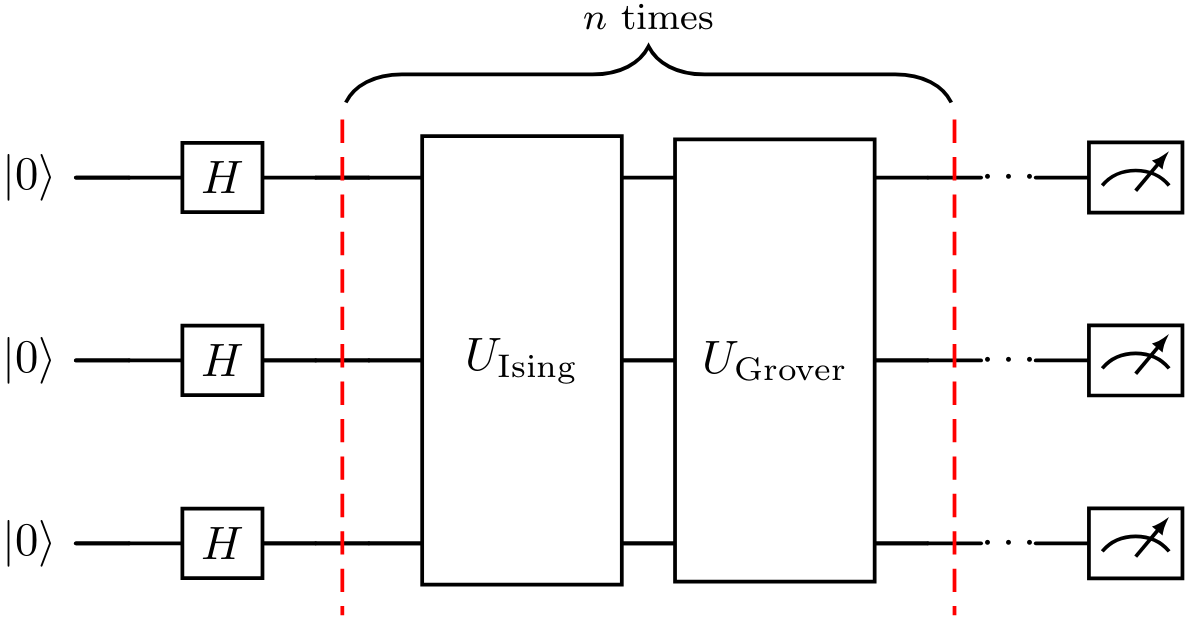}
\caption{Quantum circuit representation of the algorithm: initialization with Hadamard gates followed by one Grover iteration, where the Ising evolution operator \( U_{\text{Ising}} \) acts as the oracle, and \( U_{\text{Grover}} \) amplifies the probability of the target states.}
\label{fig:circuit}
\end{figure}

The quantum circuit for the algorithm is shown in Fig.~\ref{fig:circuit}, and it consists of the following steps:

Firstly, we initialize the qubits in the register by applying Hadamard gates to each qubit, resulting in the uniform superposition of all eigenstates of the Ising Hamiltonian:
\begin{eqnarray}
|s\rangle= \frac{1}{\sqrt{N_s}} \sum_{j=1}^{N_s}|j\rangle.
\label{Hadamard}
\end{eqnarray}

Subsequently, we perform Grover's iteration $n$ times, with each iteration consisting of two sub-steps:

(i) Application of the quantum oracle. This oracle is represented by the evolution operator for the Ising Hamiltonian corresponding to a fixed time $T$, given by:
\begin{eqnarray}
U_{\text{Ising}}=\exp (-iH_{\text{Ising}}T).
\label{Hadamard2}
\end{eqnarray}
The operator \(U_{\text{Ising}}\) can be trivially represented using \(R_z\) and \(R_{zz}\) rotation gates, and there is no need for Trotterization, as all contributions to the Hamiltonian in Eq.~(\ref{Ising}) commute with each other.  
In total, the Ising evolution requires \(\mathcal{O}(N_q^2)\) quantum gates, as the Hamiltonian includes interactions between all qubits.

Specifically, after the first application of the oracle, the state becomes:
\begin{eqnarray}
\frac{1}{\sqrt{N_s}} \sum_{j=1}^{N_s}\exp (-iE_{j}T)|j\rangle,
\label{phaseshifts}
\end{eqnarray}
where $E_j$ represents the eigenenergy corresponding to the eigenstate $|j\rangle$. Consequently, the oracle induces phase shifts that are most pronounced for the lowest energy and highest energy eigenstates.

(ii) Application of Grover's diffusion operator, defined conventionally as:
\begin{eqnarray}
U_{\text{Grover}}=2|s\rangle\langle s|-I.
\label{diffusion}
\end{eqnarray}

The role of this operator is to enhance the probability amplitudes for the lowest energy and highest energy states, where the phase shift is optimal, nearly corresponding to a phase flip akin to the usual Grover's search.

Our algorithm introduces two independent controlling parameters: $T$ and $n$, unlike the traditional Grover's algorithm characterized by a single parameter $n$. The selection of these two parameters is discussed in the subsequent subsections.

\subsection{Optimal mean evolution time \texorpdfstring{$T$}{T}} \label{s:opttime}

Let us estimate a ground state energy, which allows to evaluate the optimal evolution time.

For large values of $N_s$, the eigenenergies can be described by a probability density function given a specific set of $\varepsilon_j$ and $J_{ij}$, corresponding to a particular disorder realization. In our case, the eigenenergies follow a Gaussian distribution with a mean value of zero. The probability density function is represented as:
\begin{eqnarray}
f(E)=\frac{1}{\sigma \sqrt{2\pi}}\exp\left(-\frac{E^2}{2\sigma^2}\right),
\label{probabil}
\end{eqnarray}
where \(\sigma\) denotes the standard deviation. The model parameter \(\sigma\) 
is generally unknown and needs to be determined before executing our quantum 
algorithm. In the preprocessing stage, we randomly sample a number of bit 
strings and classically evaluate the energies corresponding to these bit 
strings with \(\mathcal{O}(N_q^2)\) computational complexity. After collecting 
\(M \gg 1\) samples, we determine the parameter \(\sigma\) with an accuracy of 
approximately \( \sim1/\sqrt{M} \). The required number of samples \(M\) is 
discussed later in Subsection~\ref{s:prob_amplitudes}.

To estimate the lowest energy $E_{\text{min}}^{*}$ and highest energy $E_{\text{max}}^{*}$  based on the probability density function in Eq. (\ref{probabil}), we solve the following equations:
\begin{eqnarray}
N_s\int_{-\infty}^{E_{\text{min}}^*} f(x)dx=1,
\label{Emin}
\end{eqnarray}
and
\begin{eqnarray}
N_s\int_{E_{\text{max}}^*}^{\infty} f(x)dx=1.
\label{Emax}
\end{eqnarray}
Introducing dimensionless quantities $e_{\text{min}}^*=E_{\text{min}}^*/ \sigma$ and $e_{\text{max}}^*=E_{\text{max}}^*/ \sigma$, we readily find that $e_{\text{max}}^*=-e_{\text{min}}^*$. Solving for $e_{\text{min}}^*$, we obtain:
\begin{eqnarray}
\frac{1}{2} \text{erfc} \frac{|e_{\text{min}}^*|}{\sqrt{2}}=2^{-N_q},
\label{erfc}
\end{eqnarray}
where $\text{erfc}$ represents the complementary error function.

The estimate of the optimal evolution time $T^*$ is derived from the condition of the phase flip for the lowest and highest energy states:
\begin{eqnarray}
|E_{\text{min}}^*| T^* = \pi,
\label{phaseflip}
\end{eqnarray}
resulting in:
\begin{eqnarray}
T^* = \frac{\pi}{\sigma} \frac{1}{|e_{\text{min}}^*|}.
\label{T}
\end{eqnarray}

For large $N_q$, an asymptotic expansion for the complementary error function in Eq. (\ref{erfc}) yields the approximate expression for $T^*$:
\begin{eqnarray}
T^* \simeq \frac{1}{\sigma} \frac{\pi}{\sqrt{2 \ln 2}} \frac{1}{\sqrt{N_q}}\left(1+\frac{1}{4\ln 2} \frac{\ln N_q}{N_q}\right).
\label{T_*}
\end{eqnarray}

Utilizing $T^*$ from Eqs. (\ref{T}) and (\ref{erfc}) allows us to enhance probability amplitudes for eigenstates $E_j$ located at the tails of the distribution, where only a countable number of eigenenergies typically reside. However, it is essential to note that $T^*$ represents the mean optimal time averaged over realizations of disorder, rather than the true optimal evolution time $T_{\text{opt}}$, which fluctuates depending on the precise values of $E_{\text{min}}$ and $E_{\text{max}}$. In this sense, $T^*$ reflects the mean optimal time $T_{\text{opt}}$ averaged over the realizations of disorder. The latter is demonstrated numerically in the subsequent section, showcasing excellent agreement between the results of numerical computation and theory. To fine-tune the optimal time and target the true lowest energy or highest energy states, methods such as Monte Carlo simulations can be employed to sample values of time in the vicinity of $T^*$ given by Eqs. (\ref{T}) and (\ref{erfc}).

It is worth noting that our approach can be extended to target states with energies $E_{\text{tar}}$ different from $E_{\text{max}}$ and $E_{\text{min}}$. This can be achieved by adopting the evolution time $\pi / |E_{\text{tar}}|$, amplifying probability amplitudes for states with energies in the vicinity of $E_{\text{tar}}$. Generally, probabilities for all states with energies $E_{\text{tar}}(2l+1)$ will be amplified, where $l$ is an integer number.  
However, since amplification occurs in the vicinity of \(E_{\text{tar}}\), the probability peaks of nearby states should not overlap significantly for the method to remain effective. The peak width is approximately \(\sim\sigma\), so choosing \(|E_{\text{tar}}| \gtrsim \sigma\) ensures that the target state remains distinguishable and is properly amplified.

\subsection{Optimal mean iteration number \texorpdfstring{$n$}{n}}

In the conventional Grover's search, the optimal number $n^*$ of iterations is determined by \cite{grover1996fast}
\begin{eqnarray}
\sin^2(2n^*+1)a \simeq 1,
\label{r}
\end{eqnarray}
where $a=\arcsin (1/\sqrt{N_s})$. For large $N_s$, this leads to $n^*\simeq \frac{\pi}{4} \sqrt{N_s}$, showcasing a quadratic speedup compared to classical search algorithms.

Numerical investigations (as detailed in the subsequent section) reveal that our proposed approach yields results consistent with Eq. (\ref{r}). Specifically, the optimal $n_{\text{opt}}$ averaged over different realizations of disorder agree well with the predictions of Eq. (\ref{r}).

\section{Main results}

\subsection{Probability amplitudes} \label{s:prob_amplitudes}

\begin{figure}[h]
\centering\includegraphics[width=0.99\textwidth]{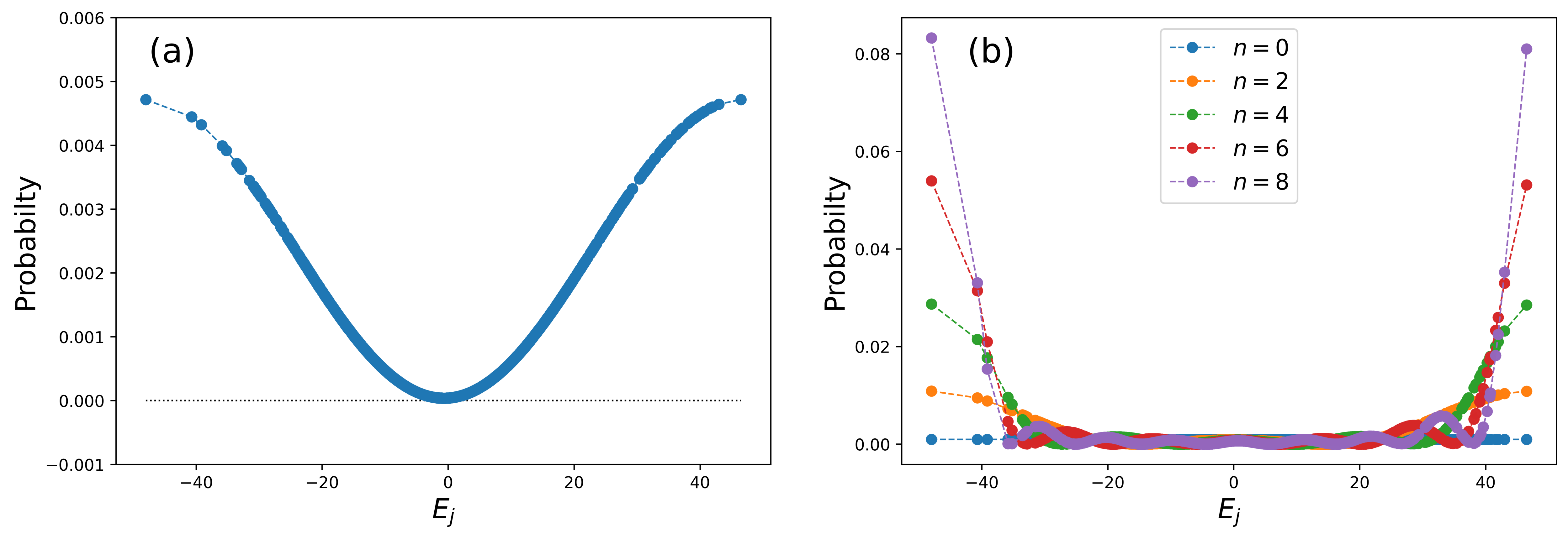}
\caption{Probability of finding the state with energy $E_j$ after the first Grover's iteration at $N_q=10$ (a); the same quantity, but after the $n$-th Grover's iteration (b). Dotted lines are guides for the eyes connecting points, which correspond to discrete eigenenergies.}
\label{fig:g}
\end{figure}

Consider the state after the first Grover's iteration, employing the quantum oracle provided by the evolution operator of the Ising model. It can be expressed as
\begin{eqnarray}
|\psi^{(1)}\rangle=\sum_{j=1}^{N_s} g_j^{(1)} |j\rangle,
\label{singleGrover}
\end{eqnarray}
where
\begin{eqnarray}
g_j^{(1)}=\frac{2}{N_s \sqrt{N_s}}\sum_{k=1}^{N_s} e^{-iE_kT} - \frac{1}{\sqrt{N_s}}e^{-iE_jT}.
\label{ampl}
\end{eqnarray}

For the disordered Ising model under consideration, in the leading order of $N_s$, the sum in the right-hand side of this equation can be replaced by an integral, yielding
\begin{eqnarray}
\sum_{k=1}^{N_s} e^{-iE_kT} \simeq N_s \int_{-\infty}^{+\infty}f(E)e^{-iET}dE = N_s e^{-\sigma^2T^2/2}.
\label{sumint}
\end{eqnarray}
Thus, we have
\begin{eqnarray}
g_j^{(1)} \simeq \frac{1}{\sqrt{N_s}}(2 e^{-\sigma^2T^2/2} - e^{-iE_jT}).
\label{ampl1}
\end{eqnarray}
The first term in the right-hand side of this equation is real, while the second term, in general, is complex. However, the second term also becomes nearly real and approaches $-1$ for $E_j$ in the vicinity of $E_{max}$ and $E_{min}$. This interplay amplifies the overall probability for the eigenenergies residing at the tails of the original distribution. In contrast, $\exp(-iE_jT)$ is nearly equal to 1 for $E_j \simeq 0$, leading to the suppression of the corresponding probability. Thus, the underlying mechanism of the algorithm can be understood in terms of positive and negative quantum interference. This scenario is valid for any distribution of eigenenergies which is symmetric with respect to the energy axis. 

By induction, it can be shown that the state after the $n$-th Grover's iteration can be represented as
\begin{eqnarray}
|\psi^{(n)}\rangle=\sum_{j=1}^{N_s} g_j^{(n)} |j\rangle,
\label{manyGrover}
\end{eqnarray}
where $g_j^{(n)}$ satisfies a recurrent relation
\begin{eqnarray}
g_j^{(n)}=\frac{2}{N_s}\sum_{k=1}^{N_s} e^{-iE_kT}g_k^{(n-1)} - e^{-iE_jT}g_j^{(n-1)},
\label{recurr}
\end{eqnarray}
with $g_j^{(0)}=1/\sqrt{N_s}$. Since at any iteration $g_j^{(n)}$ is symmetric relative to the vertical axis, the condition of probability amplification is the same as the one at $n=1$.

Figure \ref{fig:g} (a) illustrates the dependence of $|g_j^{(1)}|^2$, representing the probability of measuring the state $|j\rangle$ after the first iteration, on $E_j$ for $N_q=10$, $T=T^*$, and a specific disorder realization. To conduct these computations, the set of $E_j$ was initially computed for a given set of $\varepsilon_j$, $J_{ij}$. As observed from this figure, the probability of measuring eigenstates at the tails of the distribution is amplified, while the same probability for "inner" states is suppressed. Subsequent Grover's iterations further concentrate this distribution towards the tails of the original Gaussian distribution of eigenenergies, as depicted in Figure \ref{fig:g} (b) for $|g_j^{(n)}|^2$ and various values of $n$. These calculations were performed numerically using Eq. (\ref{recurr}) for the same disorder realization. Dotted lines are included as guides for the eyes connecting points corresponding to discrete eigenenergies.

Let us now adopt a slightly different perspective on the problem. At $k$-th step of the Grover's procedure the quantum state of the system can be presented as
\begin{equation}
    |\psi^{(n)}\rangle = a^{(n)}_\pm |\pm\rangle +\sum_{k=0}^k b_k^{(n)}|k\rangle,
    \label{eq:psik}
\end{equation}
where $|\pm\rangle$ is a superposition of two extreme energy states $|\pm\rangle = \bigl[ |+E_\mathrm{min}\rangle +|-E_\mathrm{min}\rangle\bigr]/\sqrt{2}$, $|k\rangle = \hat{U}_\mathrm{Ising}^k |0\rangle$ where
\begin{equation}
    |0\rangle = \frac1{\sqrt{N_\mathrm{s} -2}} \sum_{j \neq \pm}|E_j\rangle, 
\end{equation}
is an equal weights superposition of all non extreme states of the system.

After the Grover iteration at the optimal time duration $T=T^*$ the state $|\psi^{(n)}\rangle$ experiences  a sequence of transformations:
\begin{eqnarray}
    &&\hat{U}_\mathrm{Ising} |\psi^{(n)}\rangle = -a^{(n)}_\pm |\pm\rangle + \sum_{k=0}^n b_k^{(n)} |k+1\rangle,
    \\
    &&\hat{U}_\mathrm{Grover} \hat{U}_\mathrm{Ising} |\psi^{(n)}\rangle = a^{(n)}_\pm |\pm\rangle \Bigl[ (1-2a_0^2) |\pm\rangle + \sum_{k=0}^n b_k^{(n)} \,2a_0b_0 \langle 0| k+1\rangle\Bigr]
    \\
    &&\qquad\qquad +|0\rangle \Bigl[ -2a_0b_0 a^{(n)}_\pm + \sum_{k=0}^n b_k^{(n)}\, 2b_0^2 \langle 0| k+1\rangle \Bigr] - \sum_{k=0}^n b_k^{(n)} |k+1\rangle,
    \nonumber
\end{eqnarray}
where $a_0=\sqrt{2/N_\mathrm{s}}$, $b_0 = \sqrt{1-a_0^2}$. One can see, that during the Grover's iteration the vector $\vec{a}^{(n)} \equiv (a^{(n)}_\pm, b_0^{(n)}, b_1^{(n)}, \dots)$ composed of  the state $|\psi^{(n)}\rangle$ amplitudes, see Eq.~(\ref{eq:psik}), experiences the linear transformation: $\vec{a}^{(n)} \to \vec{a}^{(n+1)} = \hat{G}\cdot \vec{a}^{(n)}$, described by the matrix
\begin{equation}
    \hat{G} = \left[\begin{array}{cccccc}
    1-2a_0^2&  2a_0b_0 \langle 0|1\rangle& 2a_0b_0 \langle 0|2\rangle&\dots&\dots\\  
    -2a_0b_0& 2b_0^2 \langle 0|1\rangle& 2b_0^2 \langle 0|2\rangle&\dots&\dots\\
    0&-1&0&\dots&\dots\\
    0&0&-1&0&\dots\\
    \dots&\dots&\dots&\dots&\dots
    \end{array}\right].
\end{equation}

Therefore, after $n$ Grover's iterations the quantum state of the system is described by the amplitude vector $\vec{a}^{(n)} = \hat{G}^n \cdot \vec{a}^{(0)}$ of the length $n+2$ with the initial condition $\vec{a}^{(0)} = (a_0,b_0,0,\dots)$. For the disordered Ising model all statistics of the energy levels is hidden into the overlap amplitudes $\langle 0 |k\rangle$,
\begin{equation}
    \langle 0|k\rangle = \frac1{N_\mathrm{s} - 2} \sum_{j\neq \pm} e^{-ik E_j T},
\end{equation}
which at large $N_\mathrm{s}$ can be considered as an average over the energy levels statistical distribution $f(E)$,
\begin{equation}
    \langle 0 |k\rangle \approx \int dE \, f(E) e^{-ik E T}.
\end{equation}
Assuming the Gaussian distribution $f(E)$, see Eq.(\ref{probabil}), one gets the overlaps $\langle 0|k\rangle$ exponentially vanish with $k$,
\begin{equation}
    \langle 0|k\rangle = \exp\Bigl(-\frac12 k^2T^2\sigma^2\Bigr) 
    = \exp\Bigl(-\frac{\pi^2 k^2}2 \frac{\sigma^2}{E_\mathrm{min}^2}\Bigr).
\end{equation}

The computational scheme employing the $\hat{G}^n$ linear transformation of the initial amplitude vector $\vec{a}_0$ allows us calculate the dynamics of the success probability $P_\pm^{(n)} = |a_\pm^{(n)}|^2$ to find correct extreme energy state $\pm E_\mathrm{min}$ during the Grover's procedure for any statistical distribution $f(E)$ of the energy spectrum. This scheme turns out to be more efficient than a direct step by step update of the state amplitudes according to Eq. (\ref{recurr}), as it requires to keep only $\mathcal{O}(\sqrt{N_s})$ vector $\vec{a}^{(n)}$ and sparse square matrix $\hat{G}$ of the same size.

\begin{figure}[h]
\centering\includegraphics[width=0.6\textwidth]{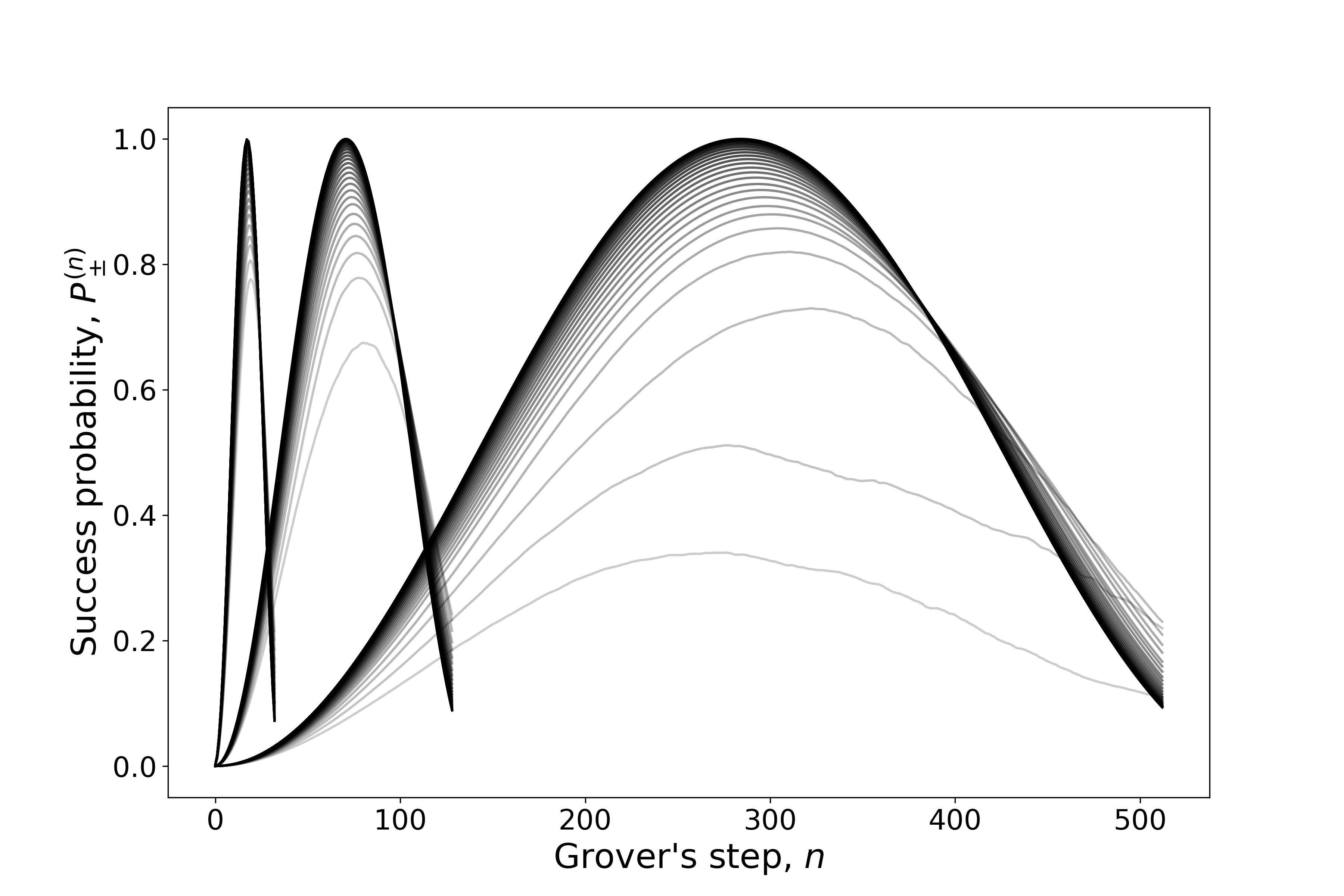}
\caption{Evolution of the success probability $P_n$ for $N_q =10, 14$ and $18$ at different levels of the disorder $\sigma$ ranging from $\sigma T = 0.01\pi$  to $\sigma T = 0.25\pi$. The disorder level is marked by the contrast with higher contrast  corresponding to a lower disorder level. Each curve was build by averaging over $400$ realizations of the energy spectrum each sampled from the normal distribution ${\cal N}(0,\sigma)$.}
\label{fig:success}
\end{figure}

To demonstrate the performance of the Grover's procedure, we plot the dynamics of the probability $P_\pm^{(n)}$ at different steps $n$ of the Grover's procedure and at different levels of disorder $\sigma$ of the Gaussian distribution $f(E)$ of the energy levels, as shown in Fig. \ref{fig:success}. It can be observed that the Grover's procedure indeed leads to an enhancement in the success probability after $n^* = \frac\pi4 \sqrt{N_s/2}$ iterations, where the factor $2$ in the square root accounts for the phase degeneracy of two extreme states after Ising evolution: $\smash{\hat{U}_\mathrm{Ising}|\pm E_\mathrm{min}\rangle = -|\pm E_\mathrm{min}\rangle}$.

The maximal success probability decreases both with the increase in disorder and the number of qubits. As the disorder increases, the probability of obtaining an energy state close to the extreme states also increases, and the Grover's procedure enhances the amplitude of this state, thereby reducing the amplitudes of the extreme states. Similarly, the same effect occurs with the increase in system size.

\begin{figure}[h]
\centering\includegraphics[width=0.6\textwidth]{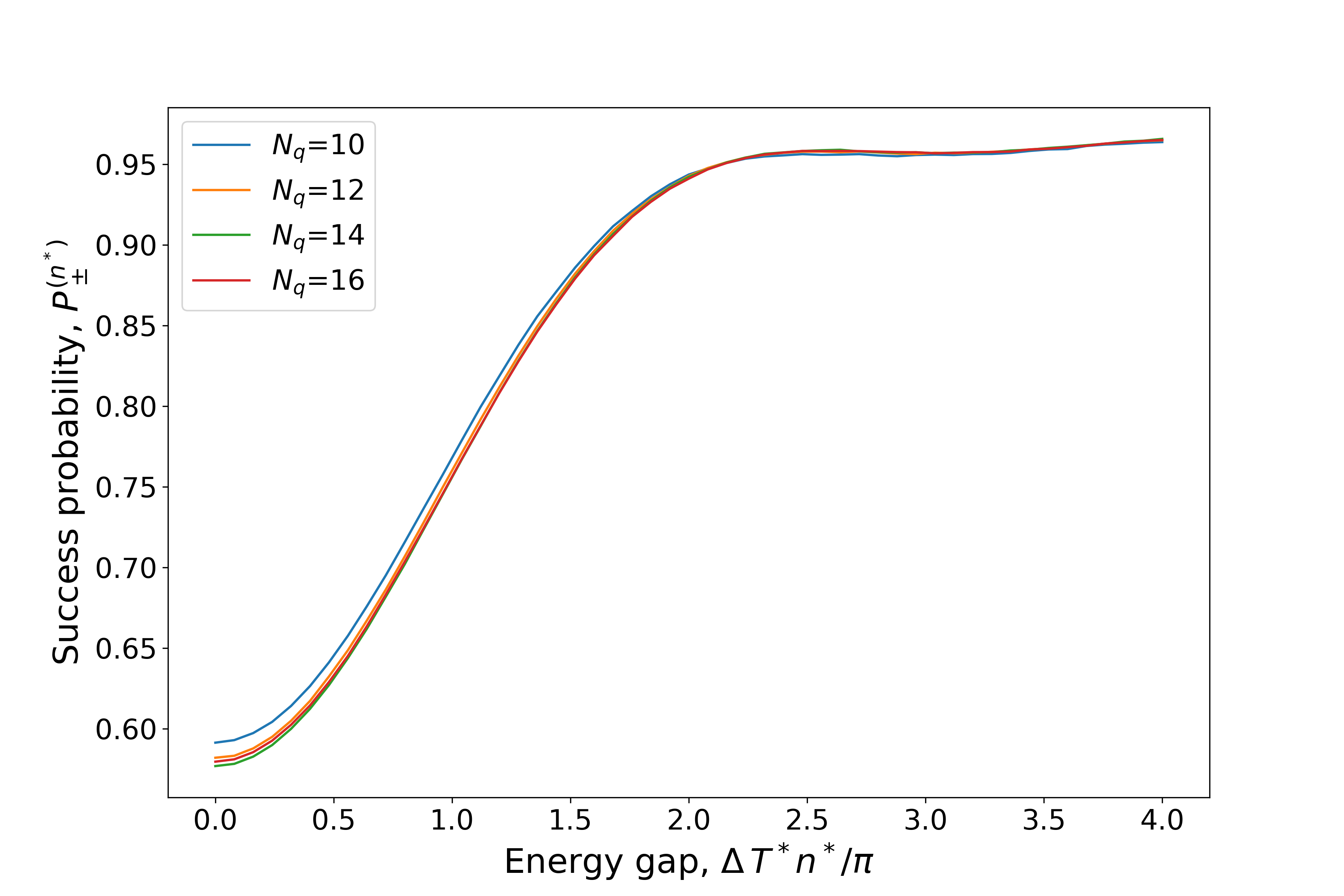}
\caption{The success probability at the optimal Grover iteration number $n^*$ as a function of the dimensionless energy gap $\delta = \Delta\, T^* n^*/\pi$ between the lowest and the next energy state at different system size.}
\label{fig:gap}
\end{figure}

In running Grover's search, we must preselect the number of Grover iterations $n^*$ under the assumption of the absence of energy states proximal to the extreme states. However, this assumption may not hold true due to our lack of knowledge regarding the actual energy state distribution. Consequently, one or more states may appear nearby. To assess the impact of closely lying states on the success probability, we sample the energy state distribution with one state positioned at an energy distance $\Delta>0$ above the minimal energy state $-|E_\mathrm{min}|$. It is observed that the success probability $P_\pm^{(n^*)}$ at the optimal Grover iteration number $n^*$ as a function of $\Delta$ saturates at $\Delta T^* n^* > 2\pi$, as illustrated in Fig. \ref{fig:gap}. Notably, in the dimensionless rescaled gap variable $\delta \equiv \Delta \, T^* n^*/\pi$, the dependence $P_\pm^{(n^*)}(\delta)$ aligns with a universal curve across all system sizes. Consequently, one may argue that the critical value of the energy gap $\Delta^*$, where the success probability becomes nonsensitive to the presence of closely lying levels, diminishes with the system size as,
\begin{equation}
    \Delta^* \sim \frac{2\pi}{T^*n^*} \propto \frac1{\sqrt{N_s}}.
    \label{eq:gapstar}
\end{equation}

\noindent
We now approximate the energy gap \( \Delta \) between the ground state and the first excited state, defined as \(\Delta = E_1 - E_0\).
Since the eigenenergies of the Ising model follow a Gaussian distribution \( f(E) \) (see Eq.~(\ref{probabil})), we estimate it as:
\begin{equation}
    \Delta \approx \sigma \left[ \Phi^{-1} \left(\frac{2}{N_s} \right) - \Phi^{-1} \left(\frac{1}{N_s} \right) \right],
    \label{eq:gap_approx}
\end{equation}
where \( \Phi^{-1}(x) \) is the quantile function (inverse cumulative distribution function), defined as
\begin{equation}
    \Phi^{-1}(x) = \inf \left\{ E \, \Bigg| \, \int_{-\infty}^{E} f(E') \, dE' \geq x \right\}.
    \label{eq:quantile_definition}
\end{equation}
Here, \(\inf\) denotes the smallest value of \(E\) for which the cumulative probability  reaches at least \( x \), effectively selecting the threshold energy corresponding to the given probability level.

For large \( N_s \), the asymptotic expansion of the quantile function is given by \cite{abramowitz1948handbook}:
\begin{equation}
    \Phi^{-1} \left(\frac{1}{N_s} \right) \approx -\sqrt{2 \ln \left( \frac{N_s}{ \sqrt{4\pi \ln N_s} } \right)}.
    \label{eq:quantile_asymptotic}
\end{equation}
Using Eq.~(\ref{eq:gap_approx}) together with the asymptotic expression for the quantile function given in Eq.~(\ref{eq:quantile_asymptotic}), we obtain the following approximation for the energy gap:
\begin{equation}
    \Delta  \approx\frac{\sigma}{\sqrt{2 \ln N_s}}.
    \label{eq:gap_estimate}
\end{equation}

\begin{figure}[h]
\centering
\includegraphics[width=0.5\textwidth]{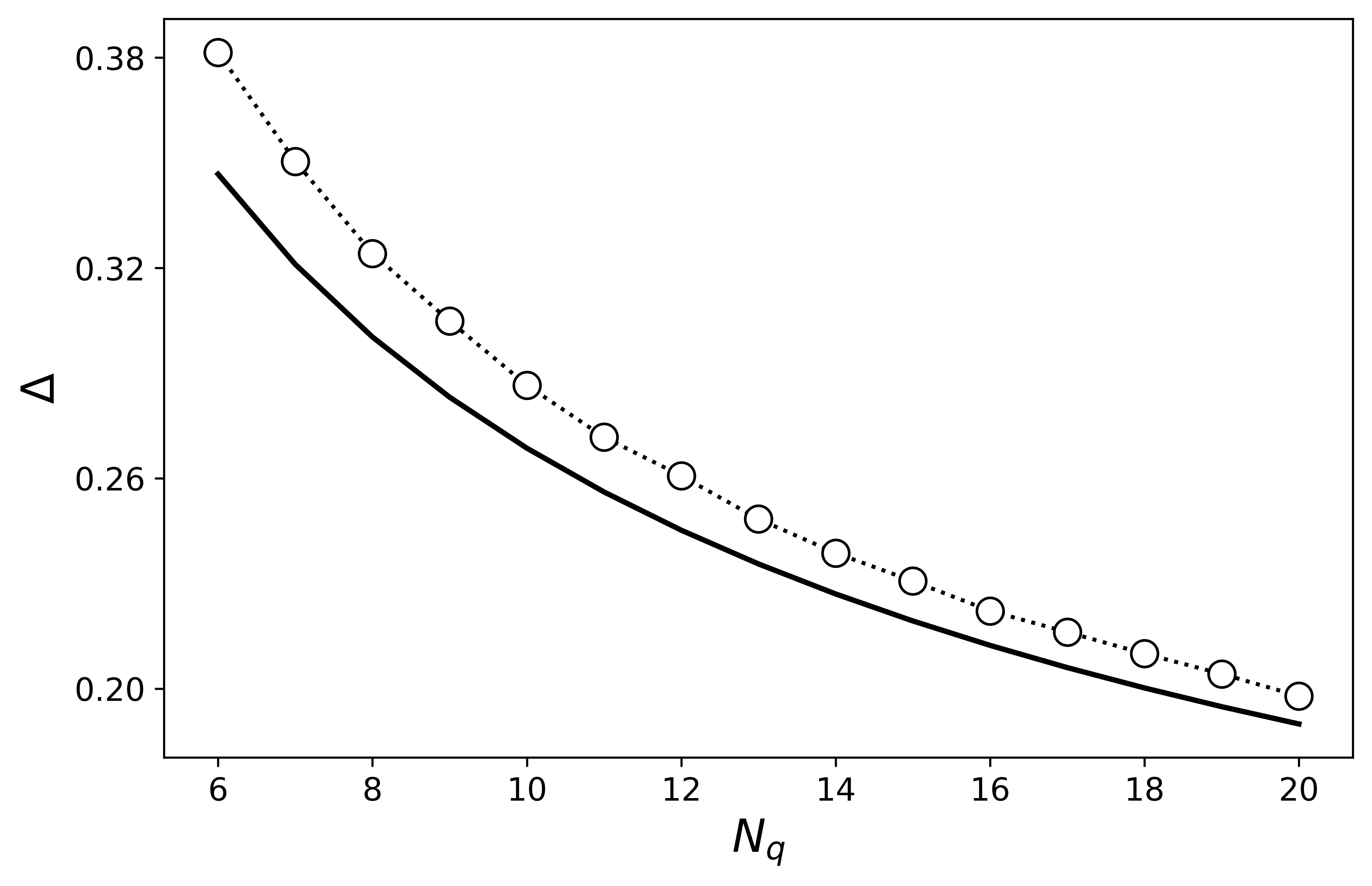}
\caption{The energy gap $\Delta $ between the ground state and the first excited state versus qubit number $N_q$: calculated numerically and averaged over $10^5$ realizations of disorder (circles), compared with the theoretical estimate from Eq.~(\ref{eq:gap_estimate}) (solid line).}
\label{fig:gap_estimate}
\end{figure}

To verify this estimate, we performed numerical simulations, generating $10^5$ disorder realizations by randomly sampling the coupling constants $J_{ij}$ and local fields $\varepsilon_j$. Fig.~\ref{fig:gap_estimate} compares the theoretical estimate from Eq.~(\ref{eq:gap_estimate}) with the numerically obtained average gap, demonstrating good agreement.

Both theoretical estimates and numerical simulations confirm the validity of our new optimization approach.

Now, it is worth discussing the relation of our results to the quantum annealing (QA) method. 
In QA, one smoothly changes the Hamiltonian of the system from an initial Hamiltonian \( H_0 \) 
to the target Hamiltonian \( H_{\text{Ising}} \) according to the function 
\(H(\lambda) = f(\lambda) H_{\text{Ising}} + (1 - f(\lambda)) H_0\)
by adiabatically varying an auxiliary parameter \(\lambda\), where \( f(0) = 0 \) 
and \( f(1) = 1 \). For a sufficiently slow rate of change \( df/d\lambda \), 
one guarantees that the system, starting in the ground state of \( H_0 \), remains in the 
ground state of \( H_{\text{Ising}} \) at the end of the evolution. 
However, during this evolution, the system may encounter a number of  avoided crossings 
between the ground state and the first excited state, with an exponentially small gap 
depending on the number of qubits. This forces an exponentially slow variation of 
the parameter \(\lambda\), which results in an overall increase in the quantum annealing 
execution time. 

Knysh \cite{knysh2016zero} analyzed such bottlenecks in the spin-glass phase 
for the Gaussian Hopfield model. He showed that the energy gap 
at these points becomes exponentially small and follows the scaling 
\(\sim\exp\left[ - c N_q^{3/4} \right]\), where $c$ is some constant (see Eq. 4 of \cite{knysh2016zero}). 
This result indicates that, in large systems, tunneling bottlenecks force quantum annealing to slow down exponentially.

In contrast, in our case, the mean energy gap remains constant during the execution of our algorithm as \(\sim1/\sqrt{N_q} \). Moreover, the number of exponentially small avoided crossings during the evolution of the parameter \(\lambda\) is not well studied and may itself be large, further complicating quantum annealing approaches.


Now, let us estimate how an uncertainty in \(\sigma\) affects the number \(n^*\) of steps in our algorithm while ensuring that it does not cause a significant change in \(n^*\). A small error in estimating the number of steps, approximately \(\delta n^* \sim 1\), is assumed to have no significant impact on efficiency.

From Eq.\~(\ref{eq:gapstar}), the variation of \(n^*\) due to the uncertainty in \(T\) is given by:
\begin{equation}
    \delta n^* \sim \frac{\delta T}{T^2 \Delta^*} \sim 1.
    \label{eq:delta_n_1}
\end{equation}
Using Eq.\ (\ref{T_*}), we estimate the error in \(T\) due to the error in \(\sigma\):
\begin{equation}
    \delta T \sim \frac{\delta \sigma}{\sigma^2} \frac{1}{\sqrt{N_q}}.
    \label{eq:delta_T}
\end{equation}
Substituting Eq.\ (\ref{eq:delta_T}) into Eq.\ (\ref{eq:delta_n_1}) gives:
\begin{equation}
    \delta n^* \sim \frac{\delta \sigma \sqrt{N_q}}{\Delta^*} \sim 1.
    \label{eq:delta_n_combined}
\end{equation}
Using Eq.\ (\ref{eq:gap_estimate}) and the previously obtained estimate for the energy gap, along with the accuracy of \(\sigma\) from Subsection \ref{s:opttime}, we obtain:
\begin{equation}
    \delta n^* \sim \frac{N_q}{\sqrt{M}} \sim 1.
    \label{eq:Nq_M_relation}
\end{equation}
Thus, the number of samples required for sufficient accuracy in \(\sigma\) scales as:
\begin{equation}
    M \sim \mathcal{O}(N_q^2).
    \label{eq:M_final}
\end{equation}
Thus, the number of required samples \(M\) grows quadratically with the number of qubits, allowing \(\sigma\) to be determined with sufficient accuracy. Even though the energy distribution consists of an exponentially large number of values, this scaling keeps the sampling process practical and efficient for estimating \(\sigma\) before running the quantum algorithm.

\subsection{Fixed evolution time and iteration number}

\begin{figure}[h]
\centering\includegraphics[width=0.99\textwidth]{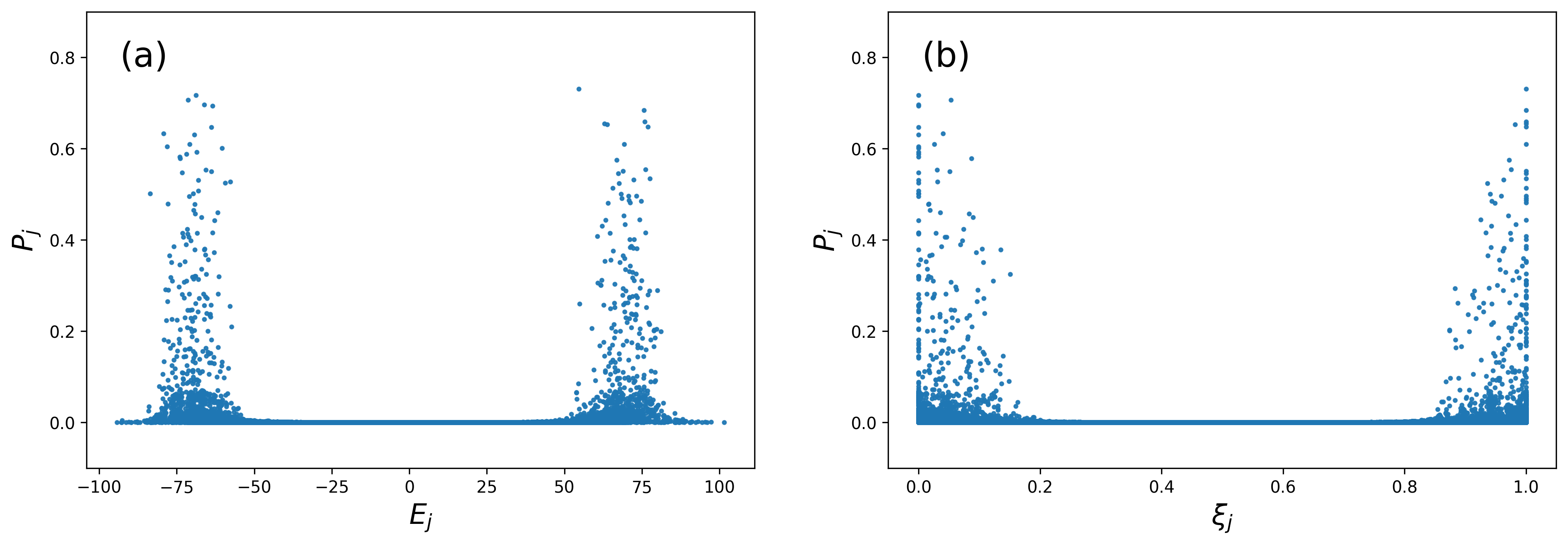}
\caption{ (a) Probability $P_j$ versus $E_j$ at $N_q=13$ for 400 realizations of disorder. (b) Probability $P_j$ versus $\xi_j$ for the same set of parameters. Both $T$ and $n$ are fixed and equal to $T^*$ and $n^*$, respectively.}
\label{fig:hist_trans}
\end{figure}

Now, we present our results obtained using a quantum-computer simulator. We use Qiskit to generate the matrix representations of the Ising evolution operator \(U_{\text{Ising}}\) [Eq.~(\ref{Hadamard2})] and the Grover diffusion operator \(U_{\text{Grover}}\) [Eq.~(\ref{diffusion})]. Instead of simulating individual gate operations, we compute the full Grover iteration operator as \(U = U_{\text{Grover}}\,U_{\text{Ising}}\) and apply \(U^n\) directly to the initial state \(\vert s\rangle\) [Eq.~(\ref{Hadamard})]. This allows us to obtain the final state vector in a single step, bypassing the need for low-level circuit emulation and significantly speeding up our calculations.


With this setup, we now examine the results of our algorithm for specific parameter choices. Let us consider the case of a fixed evolution time \(T^*\), given by Eqs.~(\ref{T}) and (\ref{erfc}), and a fixed number of Grover's iterations \(n^*\), given by Eq.~(\ref{r}). We assume a predefined set of parameters \(\varepsilon_j\) and \(J_{ij}\), which determine a certain \(\sigma\), assumed to be known. Under these conditions, we run the search algorithm with \(T = T^*\) and \(n = n^*\).
.

Firstly, we examine the probability $P_j$ of obtaining the eigenstate of $H$ corresponding to the eigenenergy $E_j$. The values of $P_j$ versus $E_j$ are depicted in Fig. \ref{fig:hist_trans} (a) for 400 different realizations of disorder at $N_q=13$. Two peaks corresponding to the tails of the initial Gaussian distribution are observed. However, the maximum energy and minimum energy states vary for different realizations of disorder, which somewhat obscures the effect of probability amplification in Fig. \ref{fig:hist_trans} (a). For this reason, in Fig. \ref{fig:hist_trans} (b), we plot the same quantity as a function of $\xi_j$, where $\xi_j$ is a realization-dependent quantity given by $\xi_j = (E_j-E_{min})/(E_{max}-E_{min})$. This quantity is exactly 0 for $E_j=E_{min}$ and exactly 1 for $E_j=E_{max}$. We observe that two peaks at 0 and 1 emerge in this plot, demonstrating the amplification of probabilities at the tails of the distribution.

\begin{figure}[h]
\centering\includegraphics[width=0.99\textwidth]{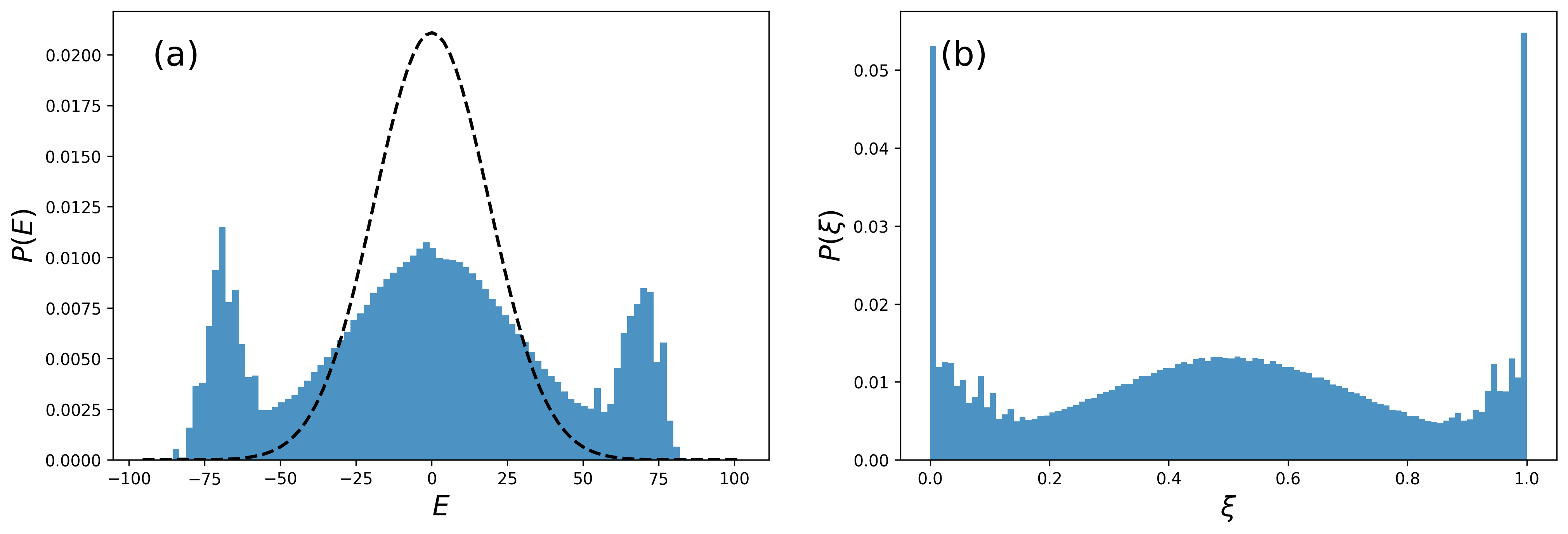}
\caption{(a) Histogram showing the probability $P^{\prime}(E)$ (blue color) at $N_q=13$, averaged over 400 realizations of disorder. For comparison, also shown is the initial distribution (dashed black line). (b) Histogram showing the probability $P^{\prime}(\xi)$ (blue color) for the same set of parameters. Both $T$ and $n$ are fixed and equal to $T^*$ and $n^*$, respectively. }
\label{fig:time2}
\end{figure}

It is also of interest to explore the same probability $P_j$ but multiplied by the density of states. This quantity, denoted as $P^{\prime}(E)$, gives the probability density function of obtaining the state with energy $E$ at the end of the algorithm. This characteristic for $N_q=13$ is shown in Fig. \ref{fig:time2} (a) in a histogram form by the blue color for 400 realizations of disorder. The initial distribution is also displayed by the dashed black line. Figure \ref{fig:time2} (b) presents the same quantity but as a function of $\xi$, where the peak structure is more pronounced.

\begin{figure}[h]
\centering\includegraphics[width=0.99\textwidth]{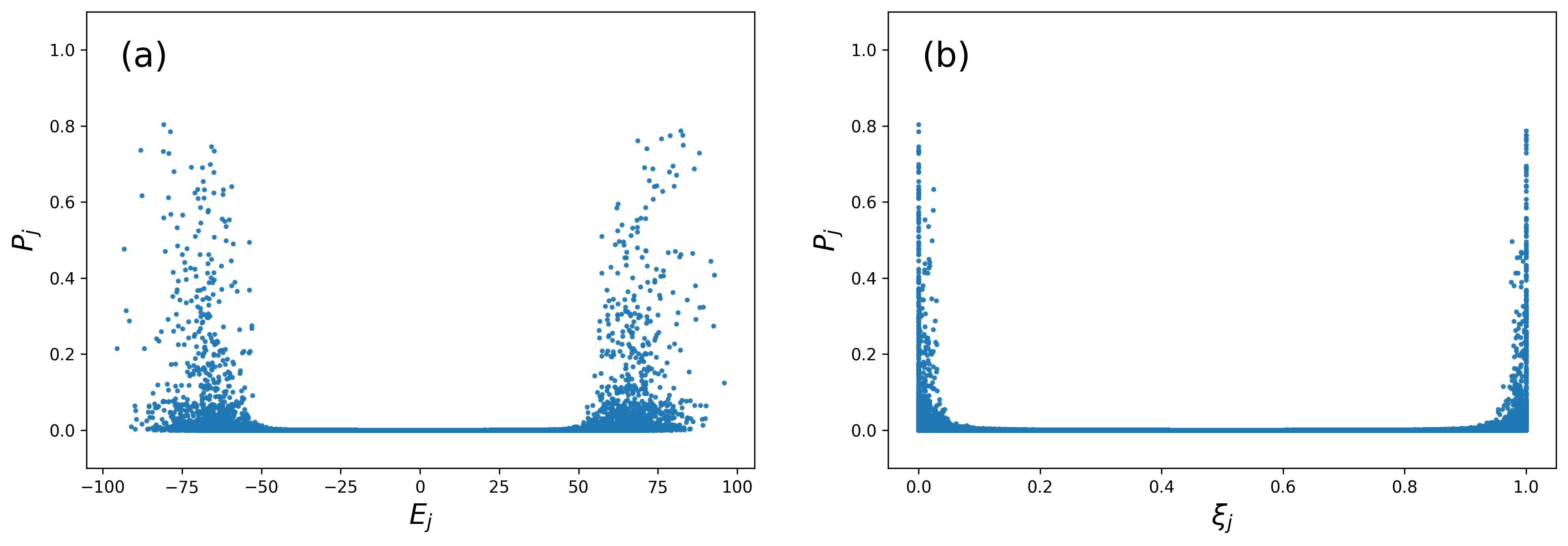}
\caption{(a) Probability $P_j$ versus $E_j$ at $N_q=13$ for 400 realizations of disorder. (b) Probability $P_j$ versus $\xi_j$ for the same set of parameters. Fine tuning of $T$ is adopted.}
\label{fig:hist_fine}
\end{figure}

\begin{figure}[h]
\centering\includegraphics[width=0.99\textwidth]{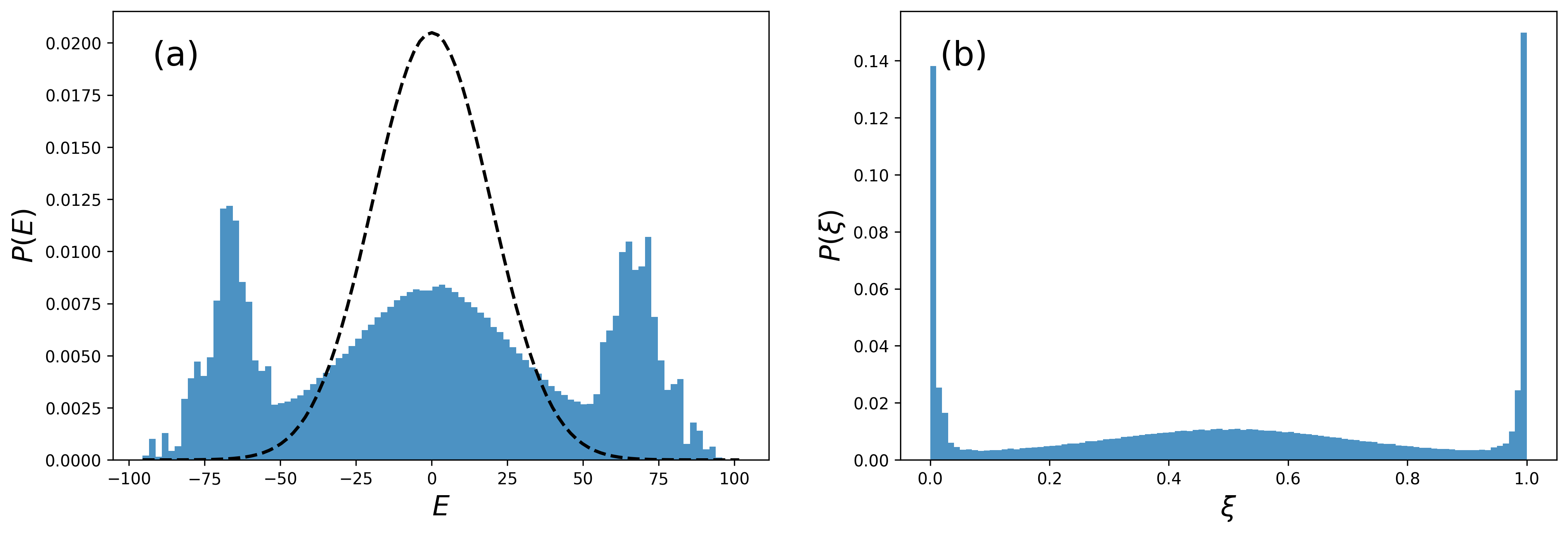}
\caption{(a) Histogram showing the probability $P^{\prime}(E)$ (blue color) at $N_q=13$, averaged over 400 realizations of disorder. For comparison, the initial distribution is also shown (dashed black line). (b) Histogram showing the probability $P^{\prime}(\xi)$ (blue color) for the same set of parameters. Evolution time $T$ is tuned for each realization of disorder, while $n=n^*$.}
\label{fig:time1}
\end{figure}


\subsection{Fine tuning of controlling parameters}
In this subsection, we explore the tuning of the main controlling parameters $T$ and $n$ depending on the realization of disorder, i.e., for different sets of $\varepsilon_j$ and $J_{ij}$ with the same normal distribution. Initially, we present our results, within classical numerical simulation, corresponding to the tuned evolution time, while $n$ is assumed to be fixed at $n=n^*$. The optimal time $T_{\text{opt}}$ can be found by adjusting $T$ in the vicinity of $T^*$, which is estimated using Eq.~\eqref{T_*}. Specifically, we define a search range $[T_a, T_b]$ around $T^*$ (e.g., with width $\sim 1/\sigma$) and discretize it into $k$ points. For illustration purposes we use $k=20$.  The algorithm is executed $k$ times, selecting the $T_i$ that maximizes the probability of the target state. Figure \ref{fig:hist_fine} (a) illustrates this effect, showing $P_j$ versus $E_j$ at $T=T_{\text{opt}}$ and $N_q=13$ for 400 realizations of disorder, while Fig. \ref{fig:hist_fine}(b) displays $P_j$ versus $\xi_j$, where the peak structure is noticeably more pronounced compared to Fig. \ref{fig:hist_trans}. The same trend holds for $P_j^{\prime}$, as shown in Fig. \ref{fig:time1} for $N_q=13$.  

Despite its simplicity, this approach clearly demonstrates the existence of an optimal evolution time \(T_{\text{opt}}\), which deviates from \(T^*\), motivating us to propose the following iterative strategy for implementation on a real quantum processor. Initially, we perform \(N_{\text{shot}} \gg 1\) executions of the algorithm with the analytically estimated evolution time \(T^*_0 = T^*\). The energies corresponding to the measured bitstrings are then computed, and the energy closest to the target energy \(E_{\text{tar}}\) is identified and denoted as \(E^*_1\). The algorithm is then re-executed with a refined evolution time \(T^*_1 = \pi / |E^*_1|\), and another set of \(N_{\text{shot}}\) measurements is collected. This procedure is iteratively repeated, updating \(T^*_n\) and refining \(E^*_n\) at each step until convergence is reached or a state matching the target energy is obtained within the desired accuracy.

\begin{figure}[h]
\centering\includegraphics[width=0.5\textwidth]{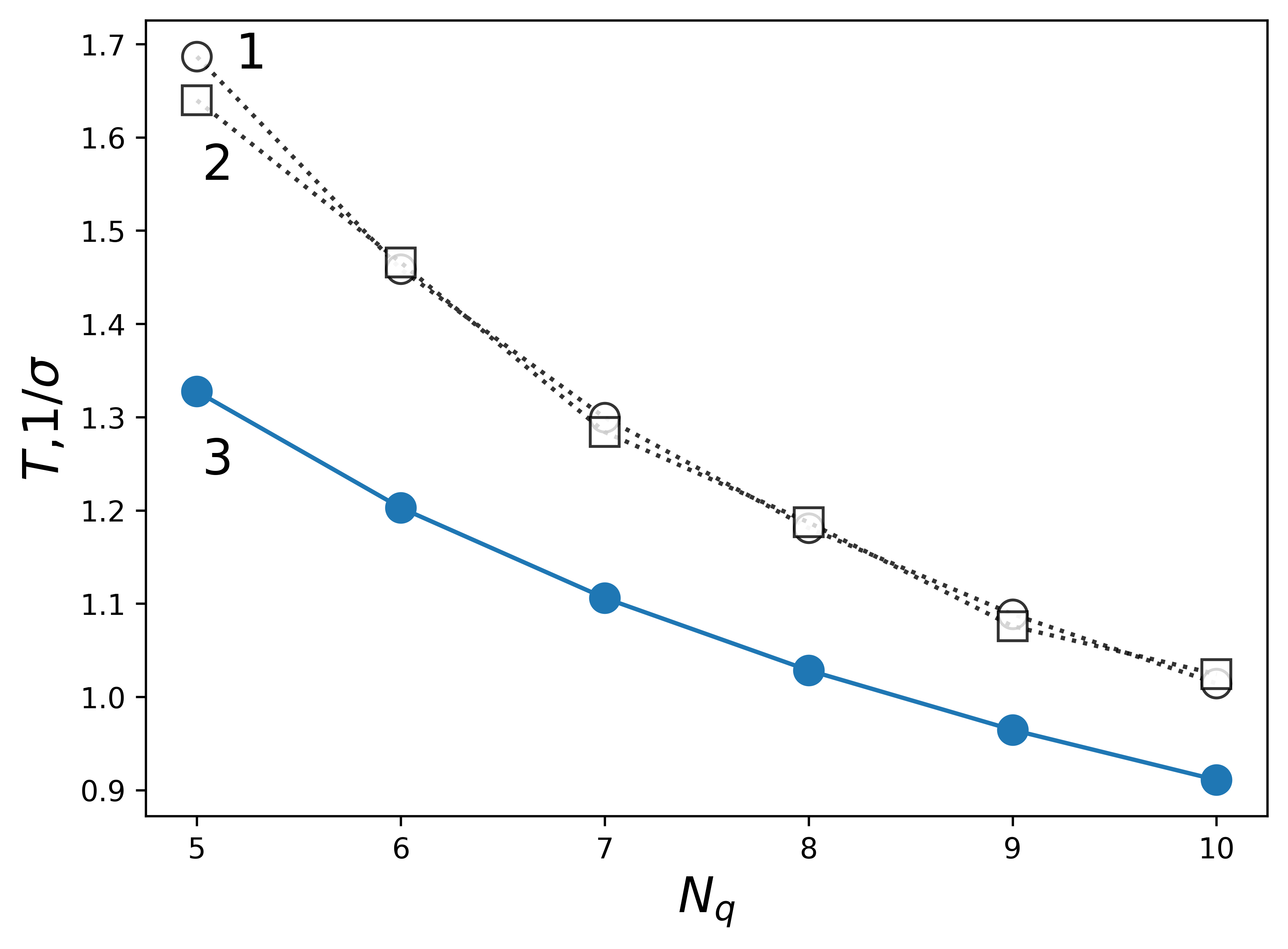}
\caption{The optimal evolution time versus qubit number $N_q$: calculated numerically and averaged over 100 realizations of disorder (curve 1), found numerically from the solution of Eq. (\ref{erfc}) (curve 2), determined by the asymptotic expansion (\ref{T_*}) (curve 3). Lines are guides for eyes connecting discrete points.}
\label{fig:opttime}
\end{figure}

\begin{figure}[h]
\centering\includegraphics[width=0.6\textwidth]{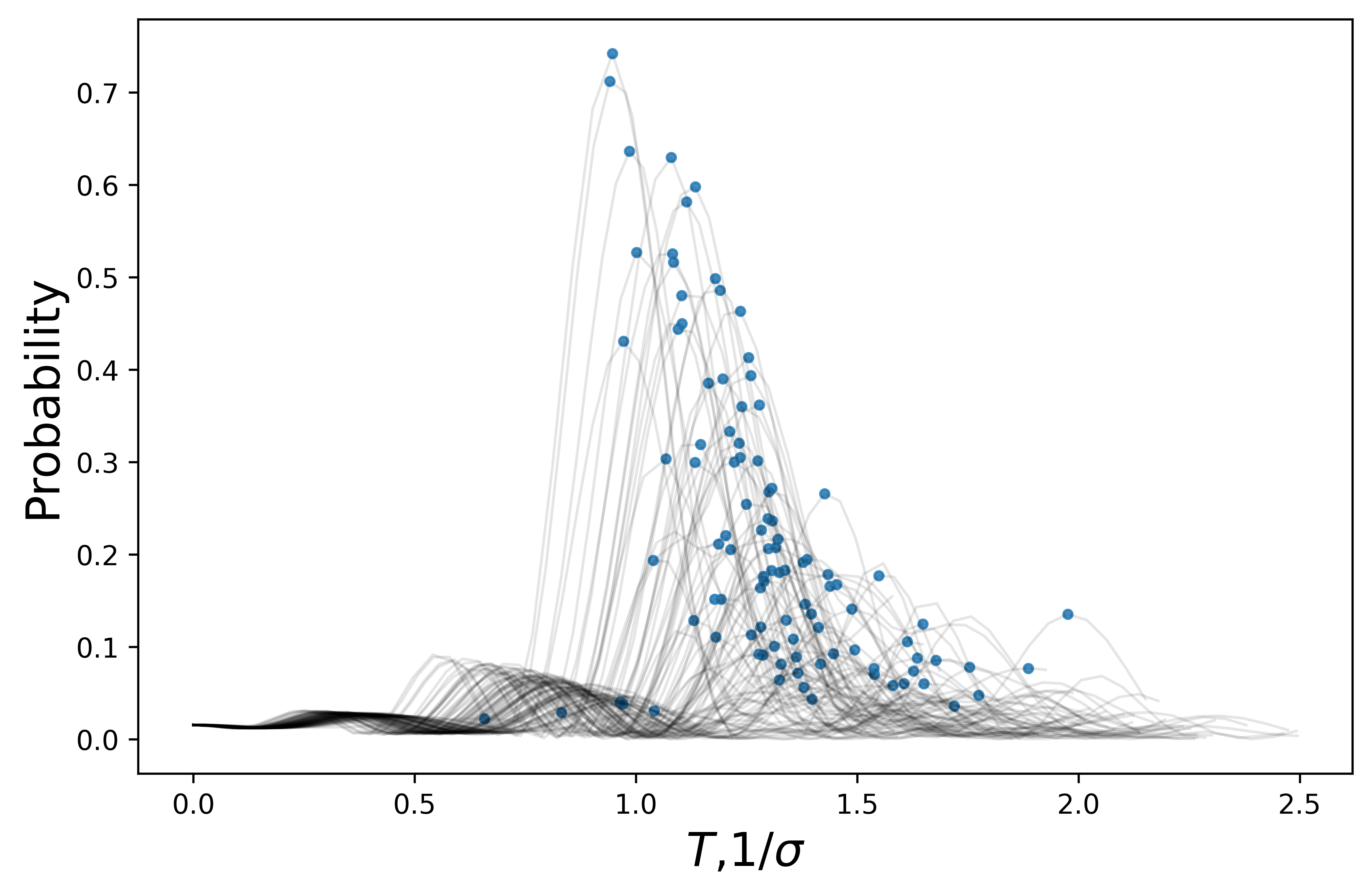}
\caption{The probability of the measuring the state with the maximum $|E_j|$ as a function of time $T$ for $N_q=6$ and 100 realizations of disorder.}
\label{fig:timeevolution}
\end{figure}

We also compare the true optimal evolution time $\overline{T}_{opt}$ averaged over realizations of disorder with $T^*$ evaluated numerically from Eqs. (\ref{erfc}) and (\ref{T_*}). The results are presented in Fig. \ref{fig:opttime} for different qubit numbers $N_q$. Lines serve as guides for the eyes, connecting discrete points corresponding to different qubit numbers. We observe an excellent agreement between the theory and the results of the numerical experiments.

\begin{figure}[h]
\centering\includegraphics[width=0.5\textwidth]{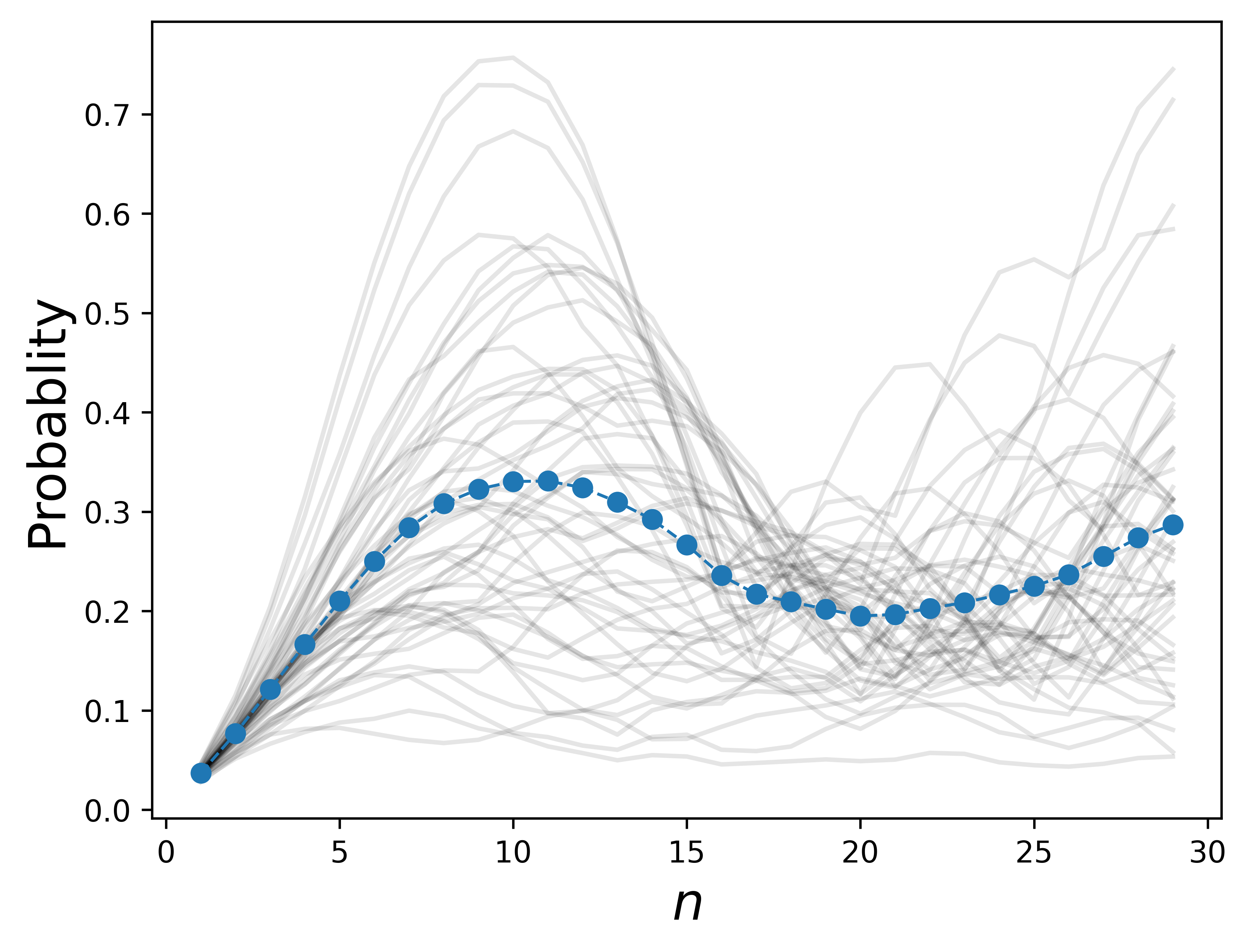}
\caption{The probability of measuring the state with the maximum $|E_j|$ at optimal time $T=T_{opt}$ for different Grover's iteration numbers $n$, $N_q=7$, and 50 realizations of disorder. Blue dots show data averaged over the realizations of disorder. Blue dashed line is guide for eyes connecting discrete points.}
\label{fig:groversteps}
\end{figure}

Figure \ref{fig:timeevolution} illustrates the dependence of the probability of obtaining the state with the maximum $|E_j|$ as a function of time $T$ for $N_q=6$ and 100 realizations of disorder. Dots indicate the maxima of these dependencies, which correspond to the true optimal evolution time $T_{opt}$. Furthermore, Fig. \ref{fig:groversteps} presents the probability of obtaining the state with the maximum $|E_j|$ measured at the optimal time $T=T_{opt}$ for different Grover's iteration numbers $n$, $N_q=7$, and 100 realizations of disorder. Blue dots represent data averaged over the realizations of disorder, with the blue dashed line serving as a guide for the eyes connecting discrete points. The first peak on this dependence corresponds to $10$ Grover's iterations, while, according to Eq. (\ref{r}), $n^*=9$ Grover's iterations are optimal. The difference between these numbers is small which demonstrates that our method operates similarly to Grover's search.

Lastly, we consider tuning not only the evolution time around $T^*$, but also tuning the Grover's iteration number around $n^*$ for each realization of disorder. This allows us to achieve realization-dependent $T=T_{opt}$ and $n=n_{opt}$. However, we found that this approach does not lead to a significant improvement in our results. This must be related to the fact that the dependence of the probability to measure the state with the maximum $|E_j|$ on $n$ is rather smooth in the vicinity of $n^*$, as expected for Grover's search.

\begin{figure}[h]
    \centering
    \includegraphics[width=0.6\textwidth]{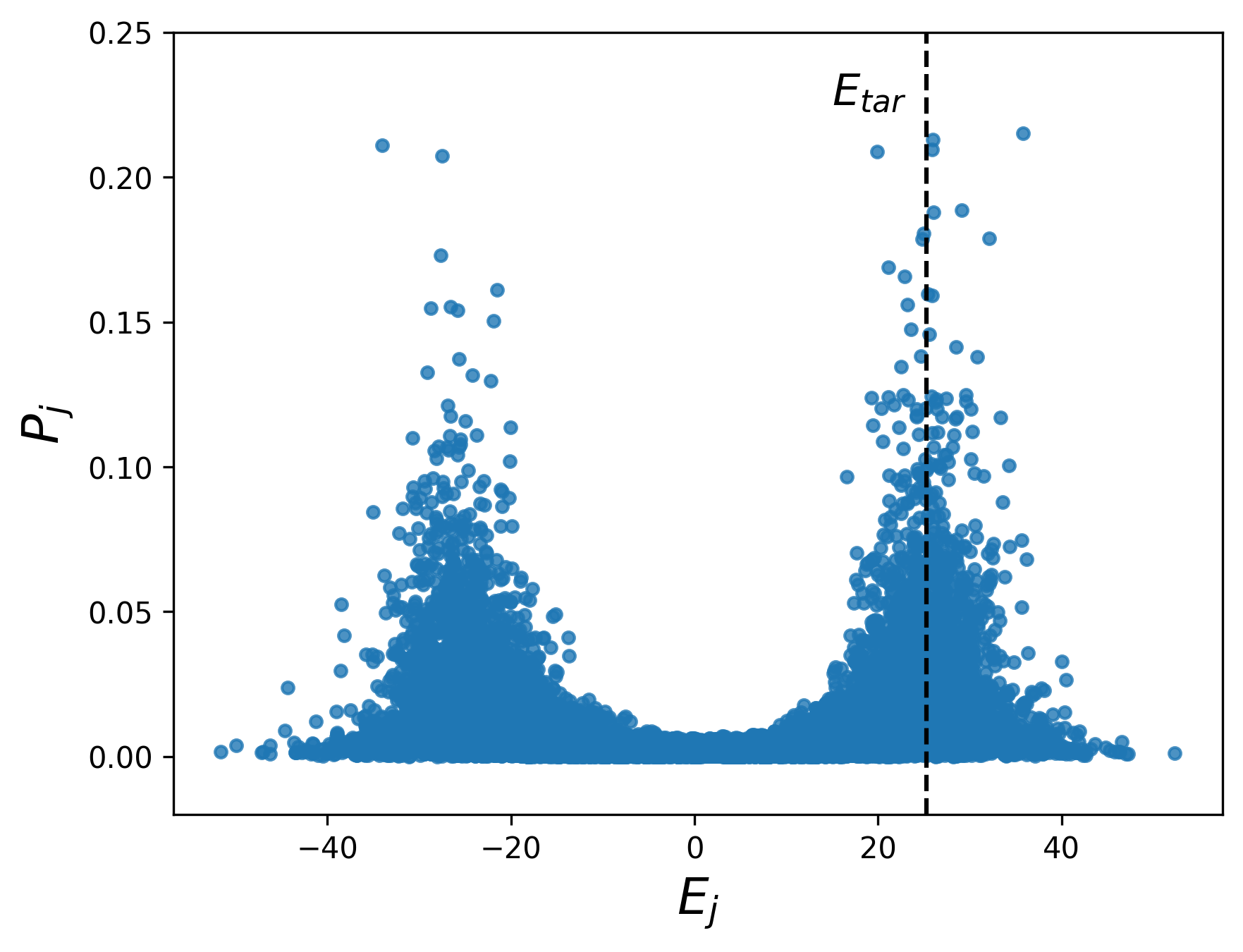}
    \caption{Probability \(P_j\) versus \(E_j\) at \(N_q = 8\) for 400 realizations of disorder. The target energy \(E_{\text{tar}}\) is set to 3 median absolute deviations from the mean energy spectrum across all realizations. The vertical dashed line marks the target state. Evolution time $T$ is tuned for each realization of disorder, while $n=n^*$.}

    \label{fig:arb2}
\end{figure}

To conclude, we examine the amplification of a target state with energy different from the extremal values. In Fig.~\ref{fig:arb2}, we demonstrate this approach for \(N_q = 8\) and 400 disorder realizations, selecting \(E_{\text{tar}}\) as 3 median absolute deviations (MAD) from the mean energy spectrum computed across all realizations. Fine tuning of \(T\) is applied to maximize the amplification of the chosen state. This method enables efficient search for higher-energy states closer to the middle of the spectrum, confirming the algorithm’s applicability beyond extremal states. 

These results show that the approach can be extended to amplify probability amplitudes for states with arbitrary energies \(E_{\text{tar}}\), provided that the resulting probability peaks remain well-separated. The peak width can be estimated as \(\sim\sigma\), meaning that \(E_{\text{tar}}\) should be sufficiently large to prevent overlap with neighboring amplified states. In practice, this condition holds when \(|E_{\text{tar}}| \gtrsim \sigma\).

\section{Conclusions}

In this paper, we have introduced a theoretical framework based on Grover's search, which can be effectively employed for the quantum simulation of Ising models and similar classical spin systems. Specifically, we have illustrated the fundamental concepts of our approach using the Ising model characterized by all-to-all interactions and random energy constants.

Central to our methodology is the utilization of the evolution operator for the Ising model as a quantum oracle within Grover's search algorithm. This operator, when applied to the uniform superposition of all eigenstates of the Hamiltonian, induces phase shifts dependent on eigenenergies and evolution time. By carefully selecting an optimal evolution time, which effectively flips the sign in front of the state with the target energy, we can enhance the probability of finding this state through Grover's search.

We have applied these principles to the task of identifying the lowest or highest energy states (or other states with closed energies) for disordered Ising models, a problem that becomes exponentially complex as the number of spins increases. Through our demonstrations, we have elucidated the fundamental operational principles of our approach.

\section*{Acknowledgments}\label{s:acknowledgments}
Useful discussions with M. A. Remnev are acknowledged.
W. V. P. and A. V. L. acknowledge support from the RSF grant No. 23-72-30004 (theoretical formalism) and from the grant of the Ministry of Science and Higher Education of the Russian Federation No. 075-15-2024-632 dated June 14, 2024 (numerical experiments). 

\section*{Data availability}
 The data used in the current study is available upon reasonable request from the corresponding authors. The code used can be found at https://zenodo.org/records/12697641.

\bibliographystyle{elsarticle-num}
\setcitestyle{square,numbers,comma}

\bibliography{neuro}


\end{document}